\documentclass{aastex}
\usepackage{spr-astr-addons}
\usepackage{graphicx}
\usepackage{txfonts}
\usepackage{xcolor}
\usepackage{tikz}
\usepackage{rotating}
\usepackage{lscape}
\usepackage{adjustbox}
\usepackage[version=4]{mhchem}
\usepackage[normalem]{ulem}
\usepackage{float}
\usepackage{natbib}
\pdfoutput=1 
\usepackage{amsmath,amstext,amssymb}
\usepackage[T1]{fontenc}
\usepackage[colorlinks=true, allcolors=blue]{hyperref}

\RequirePackage{color}
\begin{document}
	
\title{First detection of methyl formate in the hot molecular core IRAS 18566+0408}
	\shorttitle{Methyl formate towards IRAS 18566+0408}
	\shortauthors{Manna \& Pal}

	\author{Arijit Manna\altaffilmark{1}} \and \author{Sabyasachi Pal\altaffilmark{1}}
	\email{arijitmanna@mcconline.org.in}
	
	\altaffiltext{1}{Midnapore City College, Kuturia, Bhadutala, Paschim Medinipur, West Bengal, 721129, India \\email: {arijitmanna@mcconline.org.in}}

	\begin{abstract}
{  The studies of the complex molecular emission lines in the millimeter and submillimeter wavelengths towards the hot molecular cores reveal valuable
details about the chemical complexity in the interstellar medium (ISM). We presented the first detection of the rotational emission lines of the complex organic molecule methyl formate (CH$_{3}$OCHO) towards the hot
	molecular core region IRAS 18566+0408 using the high-resolution Atacama
	Large Millimeter/Submillimeter Array (ALMA) band 3 observation. The estimated
	column density of CH$_{3}$OCHO using the rotational diagram analysis was
	(4.1$\pm $0.1)$\times $10$^{15}$~cm$^{-2}$ with a rotational temperature
	of 102.8$\pm $1.2~K. The estimated fractional abundance of
	CH$_{3}$OCHO relative to hydrogen (H$_{2}$) towards the IRAS 18566+0408
	was 3.90$\times $10$^{-9}$. We noted that the estimated fractional abundance
	of CH$_{3}$OCHO is fairly consistent with the simulation value predicted
	by the three-phase warm-up model from \cite{gar13}. We also discussed the
	possible formation mechanism of CH$_{3}$OCHO towards the IRAS 18566+0408.}
	
\end{abstract}


\section{Introduction} 
 The study of the hot molecular cores (HMCs), specially in the initial evolution phases, is difficult due to large distances ($\geq$ 1 kpc), complex cluster environments, and short evolution timescales ($\leq$ 10$^{4}$ years) \citep{sil17}. The HMCs are one of the early stages of star-formation, and they play an important role in the ISM's chemical complexity \citep{shi21}. The HMCs are mainly characterised by their small source size ($\lesssim$0.1 pc), high temperature ($\gtrsim$100 K), and high gas density (n$_{\ce{H2}}\gtrsim$10$^{6}$ cm$^{-3}$) \citep{van98}. The HMC regions contain high-velocity \ce{H2O} masers that are located near Ultra-Compact (UC) H II regions \citep{meh04}. The HMCs are the most chemically rich phase in the ISM and they are characterised by the presence of complex organic molecules like methyl cyanide (\ce{CH3CN}), methyl isocyanate (\ce{CH3NCO}), methyl formate (\ce{CH3OCHO}), methanethiol (\ce{CH3SH}), dimethyl ether (\ce{CH3OCH3}), methanol (\ce{CH3OH}), ethyl cyanide (\ce{C2H5CN}), vinyl cyanide (\ce{C2H3CN}), methylamine (\ce{CH3NH2}), methylene imine (\ce{CH2NH}), aminoacetonitrile (\ce{NH2CH2CN}), and cyanamide (\ce{NH2CN}) etc \citep{sil17, beu02, shi21, hof17, gor21, ohi19, man22a, man22b}. The phase of HMCs is thought to last about $\sim$10$^{5}$ years to $\sim$10$^{6}$ years \citep{van98, gar06}.

\begin{figure*}
	\centering
	\scriptsize
	\includegraphics[width=1.0\textwidth]{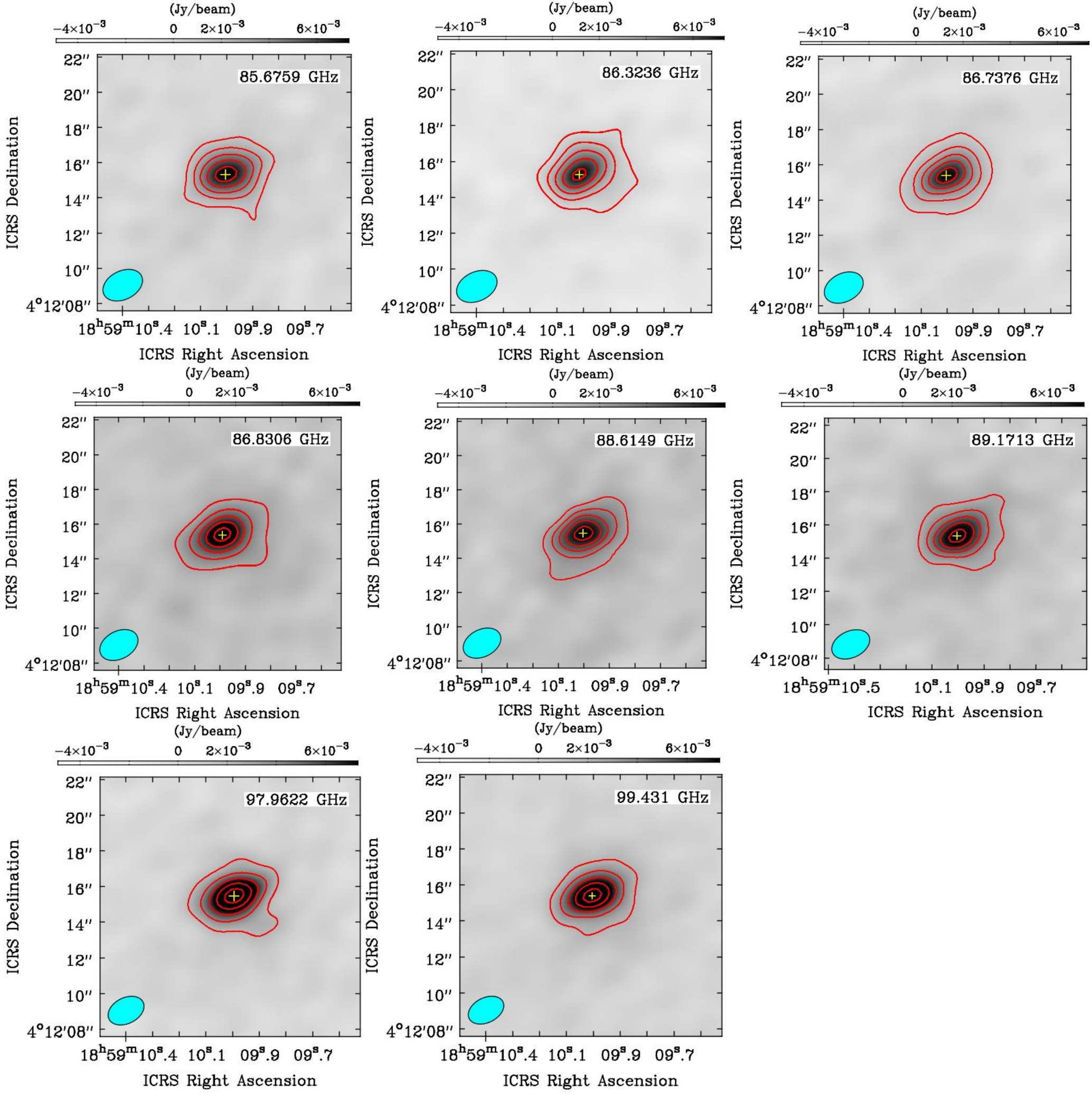}	
	\caption{Continuum images of IRAS 18566+0408 obtained with ALMA at frequency (i) 85.6759 GHz ($\sigma$ = 34.91 $\mu$Jy beam$^{-1}$), (ii) 86.3236 GHz ($\sigma$ = 43.31 $\mu$Jy beam$^{-1}$), (iii) 86.7376 GHz ($\sigma$ = 40.20 $\mu$Jy beam$^{-1}$), (iv) 86.8305 GHz ($\sigma$ = 40.75 $\mu$Jy beam$^{-1}$), (v) 88.6148 GHz ($\sigma$ = 31.93 $\mu$Jy beam$^{-1}$), (vi) 89.1716 GHz ($\sigma$ = 47.30 $\mu$Jy beam$^{-1}$), (vii) 97.9621 GHz ($\sigma$ = 45.39 $\mu$Jy beam$^{-1}$), and (viii) 99.4310 GHz ($\sigma$ = 28.99 $\mu$Jy beam$^{-1}$). The contour levels started at 3$\sigma$, where $\sigma$ is the RMS of each continuum image, and the contour levels increased by a factor of $\surd$2. The cyan circles indicate the synthesised beam of the continuum images. The corresponding synthesised beam size of all continuum images was presented in Table.~\ref{tab:cont}. The yellow cross-hair in the centre of each continuum map indicates the peak continuum position.}
	\label{fig:continuum}
\end{figure*}

 The HMC candidate IRAS 18566+0408 (alternatively, G37.55+0.20) was located at a distance of 6.7 kpc \citep{sri02}. The far-infrared luminosity of IRAS 18566+0408 was $\sim$8$\times$10$^{4}$ L$_{\odot}$, which originated from a single compact ($\geq$5$^{\prime\prime}$) dust continuum source \citep{sil17}. The single compact dust continuum emission towards IRAS 18566+0408 indicated the presence of an O8 ZAMS high-mass star \citep{zha07, sri02, sil17}. Earlier, \cite{car99} and \cite{ara05} detected the first weak radio continuum emission from IRAS 18566+0408 at wavelengths of 2 cm and 3.6 cm with a very low flux density $\sim$0.7 mJy, which indicated that this source is in the initial stage of the development of the UC H II region. Recently, \cite{hof17} detected the radio continuum emission from IRAS 18566+0408, which was resolved into 1.3 cm and 6 cm wavelengths, and they claimed that the radio continuum emission is consistent with an ionised jet. The HMC object IRAS 18566+0408 was classified as a massive disk candidate \citep{zha05}. The emission lines of maser methanol (\ce{CH3OH}) and water (H$_{2}$O) at frequencies of 6.7 GHz and 22 GHz were strongly evident towards IRAS 18566+0408 \citep{beu02}. The emission lines of maser formaldehyde (H$_{2}$CO) at wavelength 6 cm were found from the IRAS 158566+0408 \citep{ara05}. The emission lines of ammonia (NH$_{3}$) with transitions J = 1, 1 and J = 2, 2 were detected by the single-dish radio telescopes towards IRAS 18566+0408 \citep{mir94, mol96, sri02} and later \cite{zha07} studied details of the emitting region of \ce{NH3} towards IRAS 18566+0408 using the VLA.
 
\begin{table*}{}
	\centering
	\caption{Summary of the continuum images of IRAS 18566+0408.
	}
\begin{adjustbox}{width=0.8\textwidth}
	\begin{tabular}{cccccccccccc}
		\hline
		Frequency&Integrated flux&Peak flux &Beam size &Position angle& RMS\\
		(GHz)   &(mJy)    &(mJy beam$^{-1}$)&($^{\prime\prime}$$\times$$^{\prime\prime}$)&($^{\circ}$)&($\mu$Jy beam$^{-1}$)\\
		\hline
85.6759&13.53$\pm$0.66&7.16$\pm$0.24&2.399$\times$1.622&--64.029&34.91 \\
86.3236&11.49$\pm$0.94&6.58$\pm$0.36&2.382$\times$1.618&--64.527&43.31\\
86.7376&10.47$\pm$0.53&6.40$\pm$0.22&2.377$\times$1.603&--64.119&40.20\\
86.8305&13.01$\pm$0.89&7.74$\pm$0.35&2.386$\times$1.601&--64.750&40.75  \\
88.6148&12.19$\pm$0.94&6.81$\pm$0.35&2.336$\times$1.588&--64.442& 31.93  \\
89.1716&10.68$\pm$0.52&7.19$\pm$0.23&2.323$\times$1.578&--64.442&47.30   \\
97.9621&17.02$\pm$1.10&10.14$\pm$0.45&2.131$\times$1.442&--64.997& 45.39\\
99.4310&16.81$\pm$0.99&9.43$\pm$0.38&2.099$\times$1.377&--64.329&28.99\\
\hline

	\end{tabular}	
	\end{adjustbox}	
	
	\label{tab:cont}
\end{table*}

 \begin{deluxetable}{cccc}
	\tabletypesize{\footnotesize}
	\tablewidth{0pt}
	\small
	\tablecaption{Column density of hydrogen and optical depth.\label{table:column density}}
	\tablehead{\colhead{Wavelength}&\colhead{Hydrogen column density}&\colhead{Optical depth}\\
		\colhead{(mm)}&\colhead{(cm$^{-2}$)}&\colhead{($\tau_\nu$)}}
	\startdata
	3.499& $\rm{1.18\times10^{24}}$ & { 0.00495}\\
	3.472& $\rm{1.05\times10^{24}}$& { 0.00457}\\
	3.456& $\rm{1.01\times10^{24}}$ & { 0.00445}\\
	3.452& $\rm{1.21\times10^{24}}$ & { 0.00538}\\
	3.383& $\rm{9.64\times10^{23}}$ &{ 0.00473}\\
	3.361&$\rm{1.01\times10^{24}}$ &{ 0.00501}\\
	3.060&$\rm{1.06\times10^{24}}$ &{ 0.00702}\\
	3.015&$\rm{9.42\times10^{23}}$ &{ 0.00656}\\
	\hline
	{ Average Value} & $\rm{1.05\times10^{24}}$ & { 0.00533}\\
	\enddata
	
	{ }
\end{deluxetable}

 In the ISM, the complex organic molecule methyl formate (CH$_{3}$OCHO) was one of the most abundant organic molecular species, which was specially found in both high-mass and low-mass star formation regions \citep{bro75, caz03}. The emission lines of CH$_{3}$OCHO were first detected towards the Sgr B2(N) \citep{bro75}. Earlier, many chemical models indicated that the \ce{CH3OCHO} molecule was formed after the evaporation of the methanol (\ce{CH3OH}) from the grain mantle towards HMCs \citep{miller91}. \citet{gar08} presented the formation mechanism of \ce{CH3OCHO} in gas-grain interaction, but how to produce \ce{CH3OCHO} in the ISM using gas-phase reactions is not well understood. The emission lines of \ce{CH3OCHO} also detected towards another HMC candidate G31.41+0.31 with an estimated column density 3.4$\times$10$^{18}$ cm$^{-2}$ \citep{iso13}. The emission lines of CH$_{3}$OCHO were also seen in the low--mass protostar IRAS 16293--2422 \citep{caz03}. Earlier, \citet{sak06} detected the emission lines of \ce{CH3OCHO} from NGC 1333 IRAS 4B, and the author claimed these molecules can be used as a tracer of complex biomolecules in the HMC regions. Therefore, \ce{CH3OCHO} was an important molecule on the grain surfaces of hot corinos and HMCs. 

\begin{figure*}
	\centering
	\scriptsize
	\includegraphics[width=0.95\textwidth]{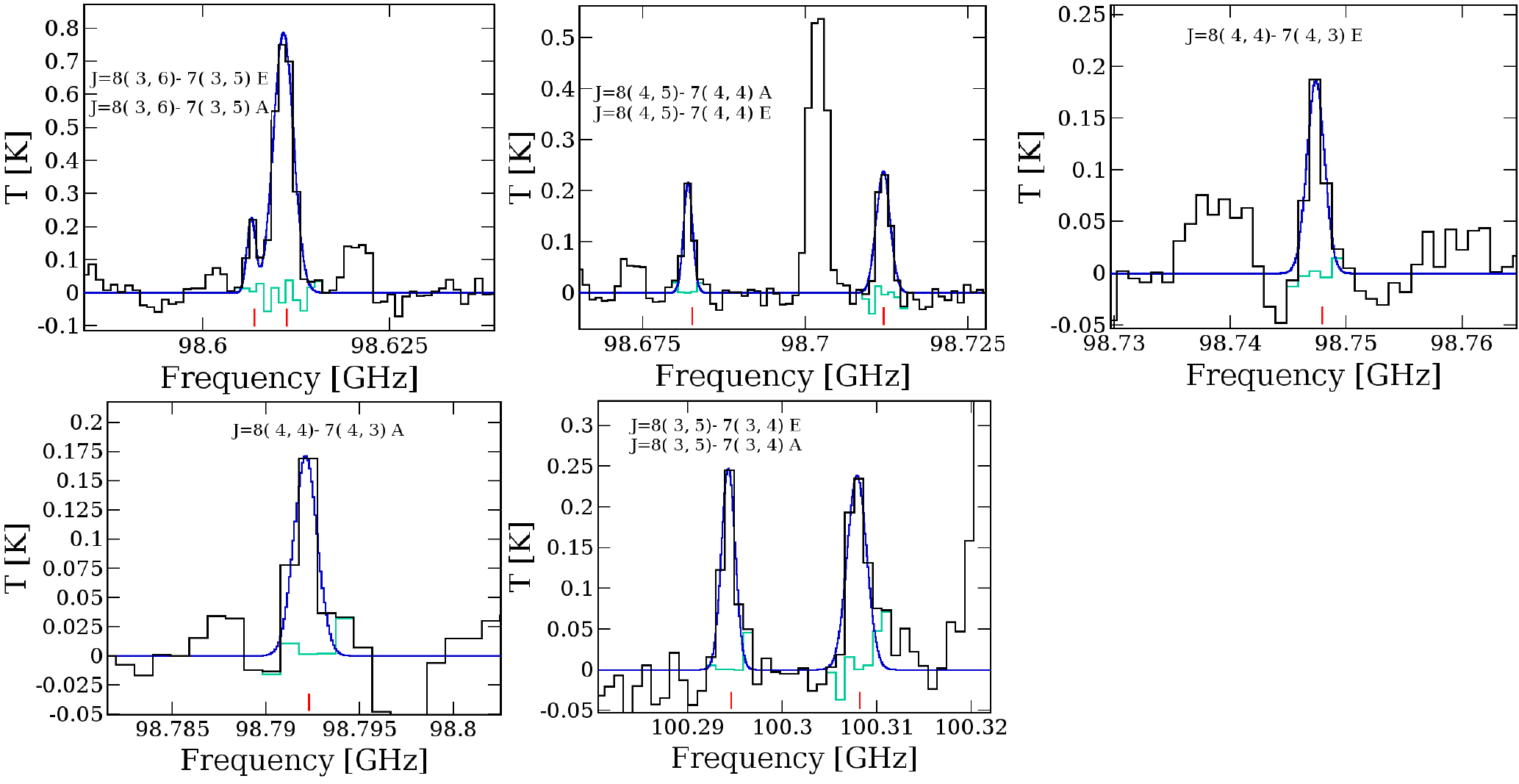}
	\caption{Rotational emission lines of CH$_{3}$OCHO between the frequency range of 85.64--100.42 GHz with their different transitions towards IRAS 18566+0408. The continuum emission has been completely subtracted from the emission spectrum. The black line represented the observed emission spectra of \ce{CH3OCHO}, while the blue line presented a Gaussian profile fitted to the observed spectra. The green line indicated the residual of the spectra.}
	\label{fig:methspec}
\end{figure*}

In this article, we presented the first detection of the rotational emission lines of \ce{CH3OCHO} towards IRAS 18566+0408 using ALMA band 3. This paper is organised as follows. In Section~\ref{obs}, we discussed the observations and data reductions. The result of the detection of \ce{CH3OCHO} was shown in Section~\ref{res}. The discussion and summary were presented in Section~\ref{con} and \ref{conclusion}.

\section{Observations and data reduction}
\label{obs}
 The millimeter-wavelength observation of HMC candidate IRAS 18566+0408 was performed with the Atacama Large Millimeter/Submillimeter Array (ALMA) using the band 3 (frequency range 85.64--100.42 GHz) receiver (Project code: 2015.1.00369.S., PI: Rosero, Viviana). The observed phase centre of IRAS 18566+0408 was $\alpha_{J2000}$: 18:59:10.000 and $\delta_{J2000}$: +04:12:16.000. During the observation, XX, YY, and XY-type signal correlators were used via the integration times of 1360.800 Sec. The observations were made on March 24th, 2016 with a minimum baseline of 41 m and a maximum baseline of 216 m. During the observation, a total of thirty-six antennas were used to study the molecular lines from IRAS 18566+0408. The solar planet Neptune was taken as a flux calibrator, J1924--2914 was taken as a bandpass calibrator, and J1830+0619 was taken as a phase calibrator. The systematic velocity ($V_{LSR}$) of IRAS 18566+0408 was $\sim$84.5 km s$^{-1}$ \citep{sil17}.

We used the Common Astronomy Software Application ({\tt CASA 5.4.1})\footnote{\url{https://casa.nrao.edu/}} for initial data reduction and spectral imaging with the standard ALMA data reduction pipeline \citep{mc07}. The continuum flux density of the flux calibrator Neptune for each baseline was scaled and matched with the Butler-JPL-Horizons 2012 flux calibrator model with 5\% accuracy using the CASA task {\tt SETJY} \citep{but12}. Initially, we calibrated the bandpass and flux by flagging the bad data using the CASA pipeline with task {\tt hifa\_flagdata} and {\tt hifa\_bandpassflag}. After the initial data reduction, we split the target data using the task {\tt MSTRANSFORM} with rest frequency in each spectral window. We used the task {\tt UVCONTSUB} for the continuum subtraction procedure from the UV plane. After the continuum subtraction, we made the spectral images of IRAS 18566+0408 using the task {\tt TCLEAN} with the rest frequency of each spectral window. After the creation of the spectral data cubes, we used the task {\tt IMPBCOR} for the correction of the primary beam pattern in the synthesised images.
\begin{figure*}
	\scriptsize
	\includegraphics[width=1.0\textwidth]{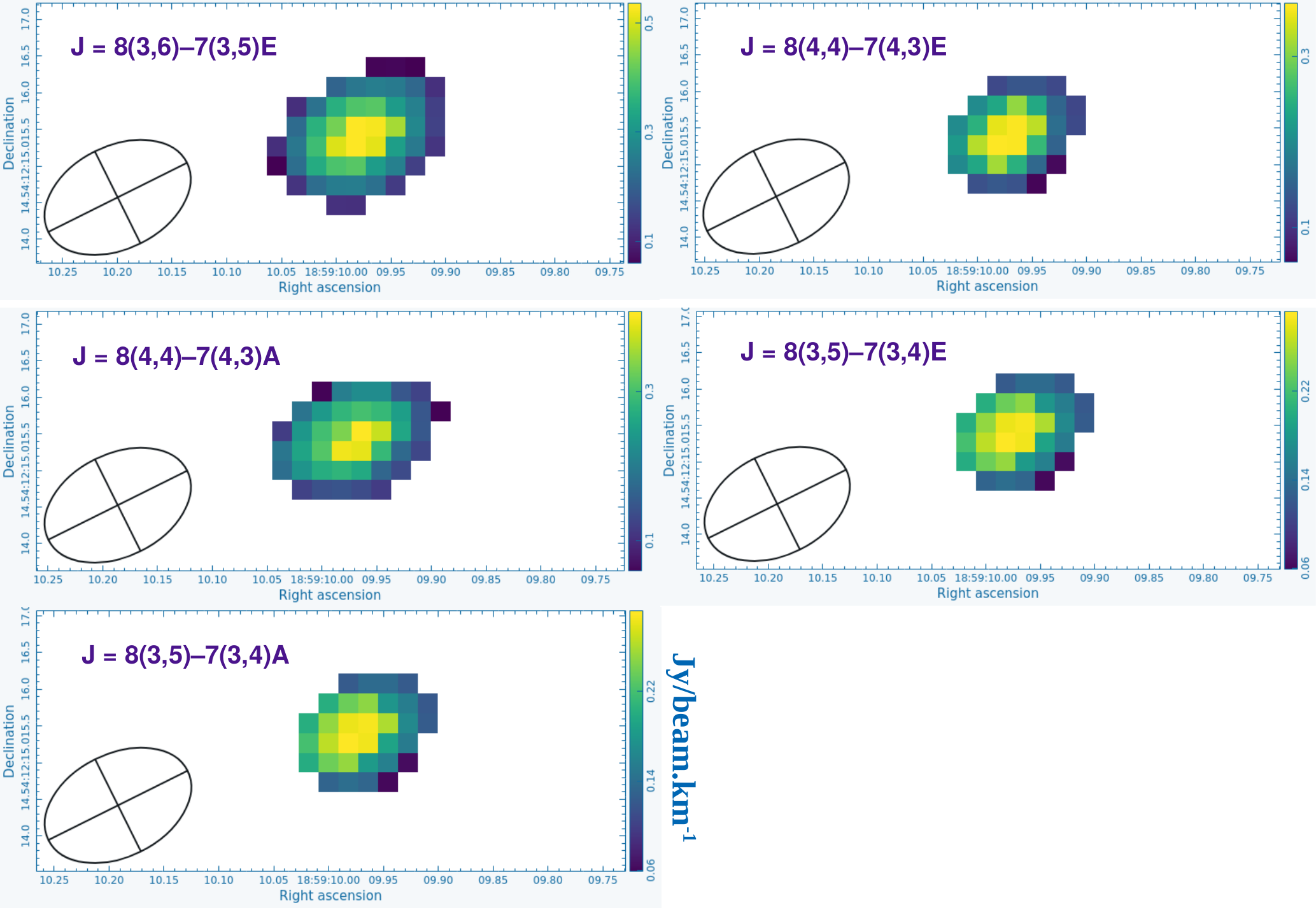}	
	\caption{Integrated emission map of unblended transitions of \ce{CH3OCHO} towards IRAS 18566+0408. The black circles indicate the synthesised beam of the integrated emission map of \ce{CH3OCHO}. The black cross indicated the major and minor axis of the synthesised beam. }
	\label{fig:map}
\end{figure*}
\section{Results}
\label{res}
\subsection{Continuum emission towards IRAS 18566+0408}
 We presented the millimeter wavelength continuum emission images towards IRAS 18566+0408 at frequencies of 85.6759 GHz (3.499 mm), 86.3236 GHz (3.472 mm), 86.7376 GHz (3.456 mm), 86.8305 GHz (3.452 mm), 88.6148 GHz (3.383 mm), 89.1716 GHz (3.361 mm), 97.9621 GHz (3.060 mm), and 99.4310 GHz (3.015 mm) in Figure~\ref{fig:continuum}, where the surface brightness colour scale has the unit of Jy beam$^{-1}$. After generating the continuum emission images, we used the CASA task {\tt IMFIT} to fit the 2D Gaussian over the continuum images and obtained integrated flux density, peak flux density, synthesised beam size, position angle, and RMS, which are presented in Table.~\ref{tab:cont}. We observed that the continuum image of IRAS 18566+0408 was larger than the synthesised beam size for each image, which indicated that the millimeter wavelength continuum emission was resolved between the wavelength 3.015 mm--3.499 mm.
 
\subsection{Estimation of hydrogen column density ($N_{\ce{H2}}$) and optical depth ($\tau_\nu$) towards IRAS 18566+0408}
 For optically thin dust continuum emission, the flux density (S$_\nu$) can be expressed as,  
 
 \begin{equation}
 S_\nu = \tau_\nu B_\nu (T_d) \Omega_{beam}
 \end{equation}
where, $\tau_\nu$ indicated the optical depth, $T_d$ is the dust temperature, ${B_\nu(T_d)}$ presented the
Planck function at dust temperature $T_d$ \citep{whi92}, and $\Omega_{beam} = (\pi/4 \ln 2)\times \theta_{major} \times \theta_{minor}$ was the solid angle of the synthesised beam. The equation of optical depth can be written as, 
 \begin{equation}
 \tau_\nu =\rho_d\kappa_\nu L
 \end{equation}
 where, $\rho_d$ presented the mass density of dust, $\kappa_{\nu}$ was the mass absorption coefficient, and $L$ indicated the path length.
The mass density of the dust can be expressed in terms of the dust-to-gas mass ratio ($Z$),
 \begin{equation}
 \rho_d = Z\mu_H\rho_{H_2}=Z\mu_HN_{H_2}2m_H/L
 \end{equation}
where, $\mu_H$ indicated the mean atomic mass per hydrogen, $\rho_{H_2}$ is the mass density of hydrogen, $m_H$ is the mass of hydrogen, and  $N_{H_2}$ is the column density of hydrogen. We used dust temperature $T_d$ = 170 K \citep{hof17}, $\mu_H = 1.41$, and $Z = 0.01$ \citep{cox00}. The estimated peak flux density of the dust continuum of the IRAS 18566+0408 at different frequencies is presented in Table.~\ref{tab:cont}. From equations 1, 2, and 3, the column density of molecular hydrogen can be expressed as, 
 
 \begin{equation}
 N_{H_2} = \frac{S_\nu /\Omega}{2\kappa_\nu B_\nu(T_d)Z\mu_H m_H}
 \end{equation}
During the estimation of the mass absorption coefficient ($\kappa_{\nu}$), we adopted the formula $\rm{\kappa_\nu = 0.90(\nu/230 GHz)^{\beta}\ cm^{2}\ g^{-1}}$ \citep{moto19}, where ${k_{230} = 0.90 \ cm^{2}\ g^{-1}}$ indicated the emissivity of the dust grains at a gas density of $\rm{10^{6}\ cm^{-3}}$, which covered by a thin ice mantle at 230 GHz. We used the dust spectral index $\beta$ $\sim$ 1.3 \citep{zha07, sil17}. Using the adopted mass absorption coefficient formula, we obtained the value $\kappa_{\nu}$ is 0.249, 0.251, 0.253, 0.254, 0.260, 0.262, 0.296, and 0.302 for the frequencies 85.6759 GHz, 86.3236 GHz, 86.7376 GHz, 86.8305 GHz, 88.6148 GHz, 89.1716 GHz, 97.9621 GHz, and 99.4310 GHz respectively. We estimated the column density of hydrogen for the eight frequency regions towards IRAS 18566+0408 which was presented in Table.~\ref{table:column density}. We take the average value to determine the resultant hydrogen column density towards IRAS 18566+0408. The obtained column density of hydrogen towards IRAS 18566+0408 was $\sim$1.05$\times$10$^{24}$ cm$^{-2}$, which was estimated by taking the average of these eight continuum values. After the estimation of the hydrogen column density, we also estimated the value of optical depth ($\tau_\nu$) using the following equation,
 \begin{equation}
T_{mb} = T_{d}(1-exp(-\tau_\nu))
\end{equation}
where, $T_{mb}$ indicated the brightness temperature and $T_{d}$ is the dust temperature of IRAS 18566+0408. During the estimation of the brightness temperature, we used the Rayleigh-Jeans approximation, 1 Jy beam$^{-1} \equiv$ 118 K. The estimated dust optical depth of eight individual frequency regions was presented in Table.~\ref{table:column density}. The average dust optical depth is estimated to be 0.00533. The estimated dust optical depth indicated that the HMC candidate IRAS 18566+0408 is optically thin between the frequency range of 85.64--100.42 GHz.
\begin{table*}{}
	\scriptsize
	\caption{Summary of the line parameters of \ce{CH3OCHO} towards IRAS 18566+0408.
	}
	\begin{adjustbox}{width=1\textwidth}
		\begin{tabular}{cccccccccccc}
			\hline 
			Species&Frequency&Transition&$E_{u}$&$A_{ij}$&Peak intensity &S${\mu}^{2}$&FWHM&V$_{LSR}$ &$\rm{\int T_{mb}dV}$ &Remark\\
			& [GHz]&[${\rm J^{'}_{K_a^{'}K_c^{'}}}$--${\rm J^{''}_{K_a^{''}K_c^{''}}}$] &[K]&[s$^{-1}$] & [K] &[Debye$^{2}$]&[km s$^{-1}$] &[km s$^{-1}$]&[K~km s$^{-1}$]&\\
			\hline
			
			CH$_{3}$OCHO&98.6069&8(3,6)--7(3,5)E&27.26&1.20$\times$10$^{-5}$&0.211&18.246&5.230$\pm$0.31&84.50$\pm$0.25&1.287$\pm$0.32&Non blended\\
			
			CH$_{3}$OCHO&98.6112&8(3,6)--7(3,5)A&27.24&1.20$\times$10$^{-5}$&0.748&18.273&7.665$\pm$0.69&84.15$\pm$0.31&7.334$\pm$0.96&Blended with \ce{C2H5CN}\\
			
			CH$_{3}$OCHO&98.6826&8(4,5)--7(4,4)A&31.89&1.05$\times$10$^{-5}$&0.212&15.966&7.658$\pm$1.23&84.21$\pm$0.55&1.358$\pm$0.29&Blended with \ce{NCCONH2}\\
			
			CH$_{3}$OCHO&98.7120&8(4,5)--7(4,4)E&31.90&1.02$\times$10$^{-5}$&0.227&15.436&7.628$\pm$0.75&84.38$\pm$0.92&1.256$\pm$0.39&Blended with \ce{NCCONH2}\\
			
			CH$_{3}$OCHO&98.7479&8(4,4)--7(4,3)E&31.91&1.02$\times$10$^{-5}$&0.187&15.440&6.682$\pm$0.98&84.18$\pm$0.86&1.287$\pm$0.21&Non blended\\
			
			CH$_{3}$OCHO&98.7923&8(4,4)--7(4,3)A&31.89&1.05$\times$10$^{-5}$&0.171&15.963&7.289$\pm$1.28&84.26$\pm$0.68&2.361$\pm$0.82&Non blended\\
			
			CH$_{3}$OCHO&100.2946&8(3,5)--7(3,4)E&27.41&1.26$\times$10$^{-5}$&0.245&18.259&6.289$\pm$0.98&84.30$\pm$0.86&8.172$\pm$1.83&Non blended\\
			
			CH$_{3}$OCHO&100.3082&8(3,5)--7(3,4)A&27.40&1.26$\times$10$^{-5}$&0.239&18.283&6.325$\pm$0.82&84.31$\pm$0.96&8.372$\pm$1.98&Non blended\\
			\hline	
			
		\end{tabular}	
	\end{adjustbox}
	
	\label{tab:LTE}
\end{table*}

\subsection{Line emission towards IRAS 18566+0408}

We extracted the millimeter wavelength spectra of IRAS 18566+0408 to create a $2.5^{''}$ diameter circular region centred at RA (J2000) = (18$^{h}$59$^{m}$09$^{s}$.92), Dec (J2000) = (4$^\circ$12$^{\prime}$15$^{\prime\prime}$.58). After the extraction of the millimeter wavelength spectra, we used the CASSIS\footnote{\url{http://cassis.irap.omp.eu/?page=cassis}} \citep{vas15} for the identification of the molecular emission lines towards IRAS 18566+0408 using the Cologne Database for Molecular Spectroscopy (CDMS)\footnote{\url{https://cdms.astro.uni-koeln.de/cgi-bin/cdmssearch}} \citep{mu05} or Jet Propulsion Laboratory (JPL)\footnote{\url{https://spec.jpl.nasa.gov/}} \citep{pic98} spectroscopic molecular databases. After the spectral analysis using the {\tt Line Analysis} module in CASSIS, we detected rotational emission lines of \ce{CH3OCHO} towards the IRAS 18566+0408 between the frequency ranges of 85.64--100.42 GHz with a spectral resolution of 1128.91 kHz. We identified a total of eight strong rotational transition lines of \ce{CH3OCHO} towards IRAS 18566+0408. There were no missing transition lines of \ce{CH3OCHO} within the observed frequency range.

{ After the identification of emission lines of \ce{CH3OCHO} from the millimeter spectra of IRAS 18566+0408, we fitted the Gaussian model over the observed spectra of \ce{CH3OCHO} using the line analysis module in CASSIS. We estimated the Full-Width Half Maximum (FWHM), quantum numbers ({${\rm J^{'}_{K_a^{'}K_c^{'}}}$--${\rm J^{''}_{K_a^{''}K_c^{''}}}$}), upper state energy ($E_u$), Einstein coefficients ($A_{ij}$), peak intensity and integrated intensity ($\rm{\int T_{mb}dV}$) after fitting a Gaussian model over the observed spectra of \ce{CH3OCHO}. The summary of the detected transitions of \ce{CH3OCHO} and Gaussian fitting parameters of \ce{CH3OCHO} was presented in Table~\ref{tab:LTE} and the observed spectra of \ce{CH3OCHO} with Gaussian fitting were shown in Figure~\ref{fig:methspec}. In the case of \ce{CH3OCHO}, the torsional substates are noticed due to the internal rotation of the methyl group. So, we observed the A and E sub-states of \ce{CH3OCHO}.  We carefully checked for the possible line contamination in the spectral profiles of \ce{CH3OCHO} with nearby molecular transitions during the spectral analysis using the line analysis module in CASSIS. We observed that J = 8(3,6)--7(3,5)A, J = 8(4,5)--7(4,4)A, and J = 8(4,5)--7(4,4)E transition lines of \ce{CH3OCHO} blended with \ce{C2H5CN} and \ce{NCCONH2} respectively. The J = 8(4,4)--7(4,3)A transition line of \ce{CH3OCHO} does not have a proper Gaussian shape due to lower spectral resolution, but this transition is not blended with other nearby molecular transitions. 

\begin{table}{}
	\centering
	\caption{Estimated emitting regions of \ce{CH3OCHO} towards IRAS 18566+0408.
	}
	\begin{tabular}{cccccccccccc}
		\hline
		Molecule&Transition&Emitting region\\
		
		&[${\rm J^{'}_{K_a^{'}K_c^{'}}}$--${\rm J^{''}_{K_a^{''}K_c^{''}}}$]&[$^{\prime\prime}$]\\
		\hline
		
		\ce{CH3OCHO}&8(3,6)--7(3,5)E&1.129\\
		&8(4,4)--7(4,3)E&1.120\\
		&8(4,4)--7(4,3)A&1.213    \\
		&8(3,5)--7(3,4)E&1.135\\
		&8(3,5)--7(3,4)A&1.133     \\
		
		\hline
	\end{tabular}

	\label{tab:prop}
\end{table}

\subsection{Spatial distribution of \ce{CH3OCHO} towards IRAS 18566+0408}
\label{source}
We created the integrated emission map of \ce{CH3OCHO} using the task {\tt IMMOMENTS} in CASA. The integrated emission maps were created by integrating the spectral data cubes in the velocity range where the emission lines of \ce{CH3OCHO} were detected. We created the emission map only for the unblended transition lines of \ce{CH3OCHO}. The integrated emission map was shown in Figure~\ref{fig:map}, which was created using the Cube Analysis and Rendering Tool for Astronomy (CARTA)\footnote{\url{https://carta.readthedocs.io/en/latest/introduction.html}} software package \citep{cm21}. The resultant emission map indicated that the emission lines of \ce{CH3OCHO} arise from the warm inner region of IRAS 18566+0408. After the generation of the emission map, we fitted a 2D Gaussian over the integrated map using CASA task {\tt IMFIT}. The deconvolved beam size of the emitting region was calculated from the following equation\\
		
\begin{equation}		
\theta_{S}=\sqrt{\theta^2_{50}-\theta^2_{beam}}		
\end{equation}
where $\theta_{50} = 2\sqrt{A/\pi}$ indicated the diameter of the circle whose area was enclosing $50\%$ line peak and $\theta_{beam}$ indicated the half-power width of the synthesised beam \citep{man22a,man22b}. The estimated emitting region of unblended transitions of \ce{CH3OCHO} is presented in Table~\ref{tab:prop}. The emitting region of \ce{CH3OCHO} varies between 1.120$^{\prime\prime}$--1.213$^{\prime\prime}$. We noticed that the emitting region of \ce{CH3OCHO} is smaller than the synthesised beam size, which indicates all transitions of \ce{CH3OCHO} were not spatially resolved.

\begin{figure*}
	\centering
	\includegraphics[width=0.48\textwidth]{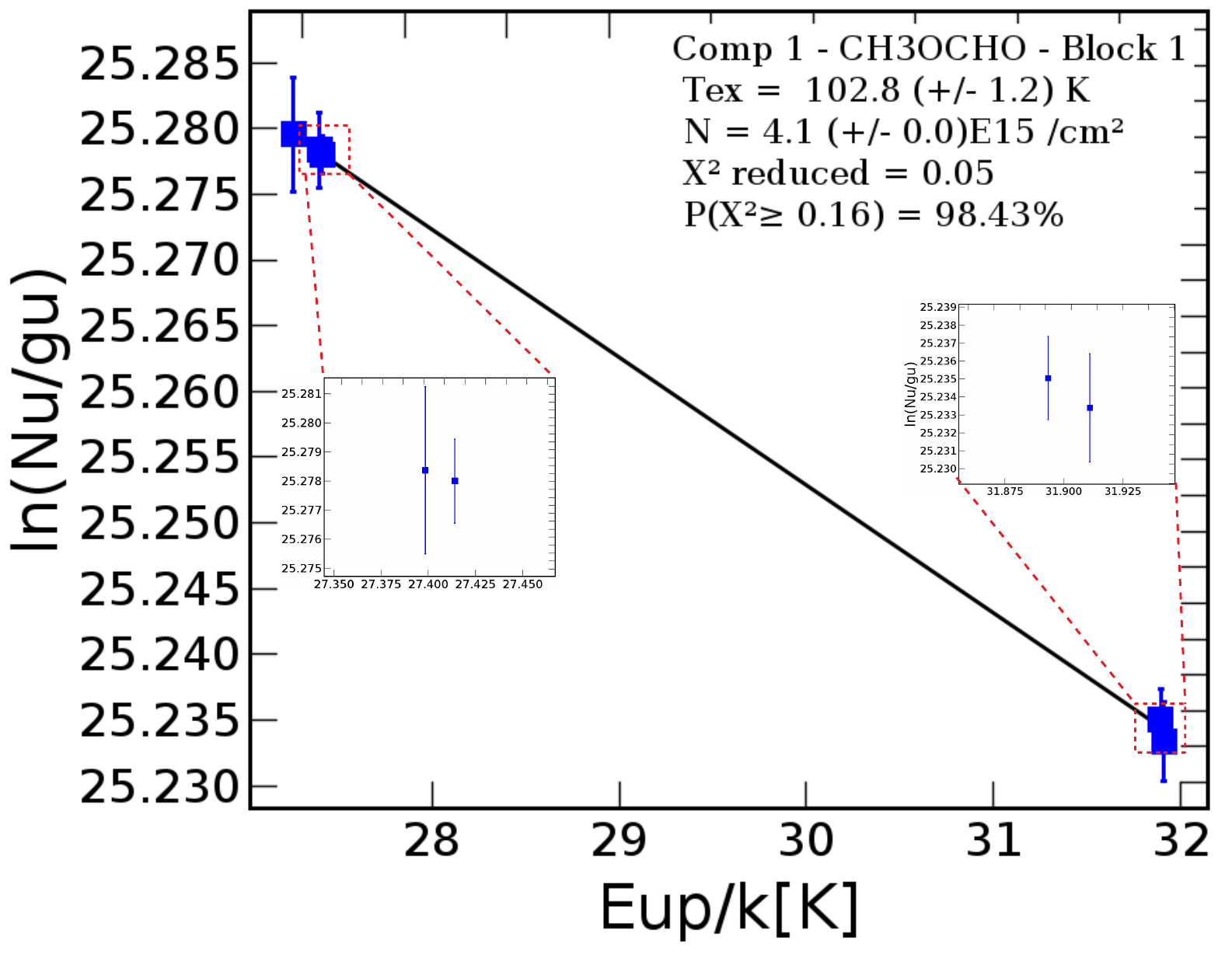}
	\caption{Rotational diagram of \ce{CH3OCHO} towards IRAS 18566+0408. The blue blocks indicated the statistical data points, and the solid black lines presented the fitted straight lines which estimated the column density and rotational temperature. The vertical blue error bars were the absolute uncertainty of $\ln(N_{u}/g_{u}$).}
	\label{fig:rotd} 
\end{figure*}

\subsection{Rotational diagram analysis}
\label{rotd}
We have detected the multiple hyperfine transitions of \ce{CH3OCHO} with different upper-state energies ($E_{u}$). So, we used the rotational diagram method to obtain the column density ($N$) in cm$^{-2}$ and rotational temperature ($T_{rot}$) in K of detected emission lines of \ce{CH3OCHO} towards IRAS 18566+0408. We used the rotational diagram method because we assumed that the observed molecular emission lines were optically thin and that they were populated in Local Thermodynamic Equilibrium (LTE) conditions. The assumption of LTE condition was reasonable towards the IRAS 18566+0408 due to very high gas density (2.6$\times$10$^{7}$ cm$^{-3}$ in the inner regions of hot core \citep{sil17}). We observed the last four unblended transitions of \ce{CH3OCHO}, i.e., J = 8(4,4)--7(4,3)E ($E_{u}$ = 31.91 K), J = 8(4,4)--7(4,3)A ($E_{u}$ = 31.89 K), J = 8(3,5)--7(3,4)E ($E_{u}$ = 27.41 K), and J = 8(3,5)--7(3,4)A ($E_{u}$ = 27.40 K) consist of two close pairs in the upper state energy ($E_{u}$) levels. These four transitions exist in the same J group but they are separated by two rotational sub-states, A and E, which occur due to internal molecular rotation of the methyl (\ce{CH3}) group \citep{sak15}. The line intensities of the A and E sublevels of \ce{CH3OCHO} are nearly similar \citep{sak15}. The rotational diagram and LTE modelling do not affect the types of \ce{CH3OCHO} transitions. The column density of optically thin molecular emission lines can be expressed as \citep{gold99}, 

\begin{equation}
{N_u^{thin}}=\frac{3{g_u}k_B\int{T_{mb}dV}}{8\pi^{3}\nu S\mu^{2}}
\end{equation}
where, $g_u$ presented the degeneracy of the upper state energy ($E_{u}$), $k_B$ is the Boltzmann constant, $\rm{\int T_{mb}dV}$ indicated the integrated intensity of the detected emission lines, $\mu$ is the electric dipole moment, $S$ is the strength of the transition lines, and $\nu$ is the rest frequency of observed molecules. Under the LTE conditions, the column density of the detected molecules can be expressed as,

\begin{equation}
\frac{N_u^{thin}}{g_u} = \frac{N_{total}}{Z(T_{rot})}\exp(-E_u/k_BT_{rot})
\end{equation}
where, $T_{rot}$ is the rotational temperature, ${Z(T_{rot})}$ is the partition function at extracted rotational temperature, and $E_u$ is the upper state energy of the observed molecules. Equation 8 can be rearranged as,
\begin{equation}
ln\left(\frac{N_u^{thin}}{g_u}\right) = ln(N)-ln(Z)-\left(\frac{E_u}{k_BT_{rot}}\right)
\end{equation}

\begin{table*}
	\caption{Comparison between the simulated and observed fractional abundance of \ce{CH3OCHO}.}\label{tab:comparison} 
	\centering      
	\begin{tabular}{c|c|c|c|cc}
		\hline 
		& \multicolumn{3}{c}{Simulated Values$^{\rm a}$} & \multicolumn{2}{c}{Observed Values$^{\rm b}$} \\
		\hline
		Species & Fast & Medium & Slow & IRAS 18566+0408\\
		\hline
		&Abundance~~~~~~~T (K)&Abundance~~~~~~~T (K)&Abundance~~~~~~T (K)&Abundance~~~~~~~~T (K)\\
		\hline
		\ce{CH3OCHO} & $9.2\times10^{-8}$~~~~120 & $3.7\times10^{-8}$~~~111 & $3.1\times10^{-9}$~~~103 & $3.90\times10^{-9}$~~~102.8 \\
		\hline
	\end{tabular}
	
	
	Notes: a -- Values taken from Table\,8 of \cite{gar13}; \\b --  this work.
	\label{tab:abundance}
\end{table*} 

Equation 9 demonstrated the linear relationship between the $E_{u}$ and $\ln (N_{u}$/g$_{u}$). The column density and rotational temperature can be estimated by fitting a straight line to the values of $\ln (N_{u}$/g$_{u}$) which is plotted as a function of $E_{u}$. The value of $N_{u}/g_{u}$ is estimated from equation 7. For the rotational diagram analysis, we extracted the line parameters like FWHM, upper state energy ($E_u$), line intensity, and integrated intensity ($\int T_{mb}dV$) using a Gaussian fitting over the observed spectra of \ce{CH3OCHO} which was presented in Table~\ref{tab:LTE}. During the rotational diagram analysis, we used only unblended transitions of the detected species. The computed rotational diagram of \ce{CH3OCHO} was shown in Figure~\ref{fig:rotd}. In the rotational diagram, the vertical blue error bars were the absolute uncertainty of $\ln (N_{u}/g_{u}$), and it was generated from the error of the observed $\int T_{mb}dV$, which was measured using the fitting of Gaussian model over observed transitions of \ce{CH3OCHO}. Using the rotational diagram analysis, we found the column density of \ce{CH3OCHO} was (4.1$\pm$0.1)$\times$10$^{15}$ cm$^{-2}$ with rotational temperature 102.8$\pm$1.2 K. The estimated rotational temperatures of \ce{CH3OCHO} are similar to the typical hot core temperature since the temperature of the hot core is above 100 K \citep{van98}. After the estimation of the column density, we estimated the fractional abundance of \ce{CH3OCHO} which was 3.90$\times$10$^{-9}$. The fractional abundance of \ce{CH3OCHO} was estimated with respect to \ce{H2}, where the column density of \ce{H2} towards IRAS 18566+0408 was 1.05$\times$10$^{24}$ cm$^{-2}$ (see Section.~{\color{blue}3.2}).

\section{Discussion}
\label{con}
Here, we presented a comparison between observational results and existing astrochemical simulations of \ce{CH3OCHO}. We followed the three-phase warm-up model of \cite{gar13} to compare the observational and simulation results. \cite{gar13} considered an isothermal collapse phase, which was followed by a static warm-up phase. In the first phase, the number density increased from $n_{H}$ = 3$\times$10$^{3}$ to 10$^{7}$ cm$^{-3}$ under the free-fall collapse, and the dust temperature was reduced to 8 K from 16 K. In the second phase, the density remained fixed at $\sim$10$^{7}$ cm$^{-3}$ where the dust temperature fluctuated from 8 K to 400 K. The temperature of IRAS 18566+0408 was $\sim$170 K, which is a typical hot core temperature, and the number density ($n_{H}$) of this source is 2.6$\times$10$^{7}$ cm$^{-3}$ \citep{hof17, sil17}. Thus, the hot core model of \cite{gar13} is suitable for understanding the chemical evolution towards the IRAS 18566+0408. \cite{gar13} used the fast, medium, and slow warm-up models based on different time scales. The time scale of the fast warm-up model is more suitable for studying the chemical evolution of HMC regions \citep{gor21}. In Table~\ref{tab:abundance}, we compare the observed fractional abundance of \ce{CH3OCHO} with the simulated abundance results of \cite{gar13} and we noticed that the slow warm-up model of \cite{gar13} satisfied our estimated abundance of \ce{CH3OCHO} towards IRAS 18566+0408. \citet{gar13} estimated that the abundance of \ce{CH3OCHO} towards the HMCs environment is 3.1$\times$10$^{-9}$ with temperature 103 K under the slow warm-up conditions and we estimated the abundance of \ce{CH3OCHO} towards HMC object IRAS 18566+0408 was 3.90$\times$10$^{-9}$ with temperature 102.8 K, which indicated a good agreement between the simulation and observational results.

{In the ISM, the \ce{CH3OCHO} molecule can be efficiently created on the surface of dust grains via the reaction between methoxy radical (\ce{CH3O}) and formyl radical (HCO) (HCO+\ce{CH3O}$\rightarrow$\ce{CH3OCHO}). This reaction is the main formation route in the three-phase warm-up chemical model by \citet{gar13}}. The chemical simulation between \ce{CH3O} and HCO shows that these radicals are mobile around 30--40 K and the reaction is efficient for the formation of \ce{CH3OCHO} in the HMCs \citep{gar13}. According to the simulation of Figure 1 in \citet{gar13}, it is clear that the gas phase \ce{CH3OCHO} in the HMC region mainly comes from the ice phase. Our estimated abundance of \ce{CH3OCHO} (3.90$\times$10$^{-9}$) towards IRAS 18566+0408 is similar to the simulated abundance of \ce{CH3OCHO} (3.91$\times$10$^{-9}$) under the slow warm-up model of \citet{gar13}, which indicated that \ce{CH3O} and HCO are responsible for the production of \ce{CH3OCHO} on the surface of dust grains in IRAS 18566+0408. The slow-warm phase of \ce{CH3OCHO} indicated that a significant amount of \ce{CH3OCHO} is destroyed due to the evaporation of the methanol (\ce{CH3OH}) in the hot core region of IRAS 18566+0408.

\section{Summary}
\label{conclusion}
In this article, we presented the detection of complex biomolecule \ce{CH3OCHO} at millimeter wavelengths towards the HMC candidate IRAS 18566+0408 using the ALMA. The main results are as follows.\\
$\bullet$ We reported the first detection of the rotational emission lines of \ce{CH3OCHO} towards the HMC region IRAS 18566+0408 using ALMA band 3 between the frequency range of 85.64--100.42 GHz.\\
$\bullet$ From the dust continuum emission, the estimated column density of hydrogen ($N_{H_2}$) was 1.05$\times$10$^{24}$ cm$^{-2}$. We also estimated that the dust optical depth was 0.00533, which indicated IRAS 18566+0408 is optically thin between the frequency range of 85.64--100.42 GHz.\\
$\bullet$ The estimated column density of \ce{CH3OCHO} towards IRAS 18566+0408 was (4.1$\pm$0.1)$\times$10$^{15}$ cm$^{-2}$ with rotational temperature 102.8$\pm$1.2 K. The estimated fractional abundance of \ce{CH3OCHO} towards IRAS 18566+0408 with respect to \ce{H2} is 3.90$\times$10$^{-9}$. Our estimated rotational temperature indicated that the emission lines of \ce{CH3OCHO} arise from the warm inner region of IRAS 18566+0408 because the temperature of the hot core, in general, is above 100 K.\\
$\bullet$ We compared our estimated abundance of \ce{CH3OCHO} with the three-phase warm-up model of \cite{gar13}. After the comparison, we noticed that the slow warm-up model of \cite{gar13} satisfied the abundance of \ce{CH3OCHO} towards IRAS 18566+0408.\\
$\bullet$ After the successful detection of \ce{CH3OCHO} towards IRAS 18566+0408, a broader study was needed to search for other molecular lines in the other frequency bands of ALMA to understand the chemical complexity in this HMC.}

\section*{Acknowledgments}{We thank the anonymous referee for the helpful comments that improved the manuscript. This paper makes use of the following ALMA data: ADS /JAO.ALMA\#2015.1.00369.S (PI: Rosero, Viviana). ALMA is a partnership of ESO (representing its member states), NSF (USA), and NINS (Japan), together with NRC (Canada), MOST and ASIAA (Taiwan), and KASI (Republic of Korea), in co-operation with the Republic of Chile. The Joint ALMA Observatory is operated by ESO, AUI/NRAO, and NAOJ. }\\\\

\section*{Data availability}{The data that support the plots within this paper and other findings of this study are available from the corresponding author upon reasonable request. The raw ALMA data are publicly available at \url{https://almascience.nao.ac.jp/asax/} (project id:  2015.1.00369.S).} 

\section*{Funding} No funds or grants were received during the preparation of this manuscript.

\section*{Conflicts of interest}
The authors declare no conflict of interest.

\section*{Author Contributions}
S.P. conceptualize the project. A.M. analysed the ALMA data and identify the emission lines of methyl formate (\ce{CH3OCHO}) from HMC candidate IRAS 18566+0408. A.M analyses the rotational diagram to derive the column density and rotational temperature of \ce{CH3OCHO}. A.M. and S.P. wrote the main manuscript text. All authors reviewed the manuscript.

\makeatletter
\let\clear@thebibliography@page=\relax
\makeatother

\end{document}